\documentclass[prc,twocolumn,showpacs,superscriptaddress,floatfix,nofootinbib]{revtex4}
\usepackage{amsmath}
\usepackage{graphicx}
\pdfoutput=1

\def\be{\begin{equation}}
\def\ee{\end{equation}}
\def\ba{\begin{eqnarray}}
\def\ea{\end{eqnarray}}
\def\bas{\begin{eqnarray*}}
\def\eas{\end{eqnarray*}}

\begin{document}

\title{Occupation number-based energy functional for nuclear masses}

\author{M. Bertolli} 

\affiliation{Department of Physics and Astronomy, University of
Tennessee, Knoxville, TN 37996, USA}

\author{T. Papenbrock} 

\affiliation{Department of Physics and Astronomy, University of
Tennessee, Knoxville, TN 37996, USA}

\affiliation{Physics Division, Oak Ridge National Laboratory, Oak
Ridge, TN 37831, USA}

\author{S.M. Wild}

\affiliation{Mathematics and Computer Science Division, Argonne National
Laboratory, Argonne, IL 60439, USA}

\date{\today}

\begin{abstract}
  We develop an energy functional with shell-model occupations as the
  relevant degrees of freedom and compute nuclear masses across the
  nuclear chart. The functional is based on Hohenberg-Kohn theory with
  phenomenologically motivated terms.  A global fit of the
  17-parameter functional to nuclear masses yields a root-mean-square
  deviation of $\chi = 1.31$ MeV.  Nuclear radii are computed
  within a model that employs the resulting occupation numbers.
\end{abstract}


\maketitle

\section{Introduction}

Nuclear theory seeks to formulate a consistent framework for nuclear
structure and reactions.  Nuclear density functional theory (DFT) is a
particularly useful tool for describing the ground-state properties of
nuclei across the nuclear chart~\cite{Gor02,Ben03,Lun03,Sto03}.  This
approach is based on the pioneering works of Skyrme~\cite{Sk56} and
Vautherin~\cite{Vau72,Vau73} and might theoretically be based on the
theorems by Hohenberg and Kohn~\cite{Hoh64}.  It has been applied
through the self-consistent mean-field computations with
density-dependent energy functionals~\cite{Koh65}.  Formally, the
energy-density functional results from the Legendre transform of the
ground-state energy as a functional of external one-body
potentials~\cite{Lieb83}.  While DFT allows for a simple solution to
the quantum many-body problem, the construction of the functional
itself poses a challenge~\cite{unedf,unedf2}. Various efforts aim at
constraining the functional from microscopic
interactions~\cite{stoitsov} or devising a systematic
approach~\cite{carlsson}.  In this paper we explore the development of
an energy functional that employs shell-model occupations instead of
the nuclear density. The minimization of this functional is
technically simpler than DFT, and global mass-table fits of the
functional require only modest computational resources.
	
In nuclear physics, energy-density functional theory is a practical
tool that is popular because of its computational simplicity and
success~\cite{Ben03,Irina}.  The universality of the functional (i.e., 
the possibility to study nucleons in external potentials) is seldom
used~\cite{neutrondrops}. This distinguishes nuclear DFT from DFT in
Coulomb systems and makes alternative formulations worth studying. For
computational simplicity, we would like to maintain the framework of
an energy functional. However, there is no need to focus only on
functionals of the density.  The description and interpretation of
nuclear systems are often based on shell-model occupation numbers
rather than densities~\cite{MayerJensen}.  For instance, in nuclear
structure one is often as much interested in the occupation of a given
shell-model orbital or the isospin-dependence of the effective
single-particle energies as in the shape of the density distribution.
This approach is natural for shell-model Hamiltonians that are based
on single-particle orbitals.  The purpose of this paper is to
develop a nuclear energy functional constructed in a shell-model
framework.  We show that an occupation number-based functional can
be employed in global mass-table computations and can 
perform at a competitive level, with a reasonable number of
parameters and relatively short computation time.
	
Many approaches to the nuclear energy-density functional are
empirical~\cite{Dob84,Oli88,Rob01,Dug01}, but guidance can be found
from analytical solutions to simpler models.  The complete
ground-state energy of a quantum many-body system as a functional of
any external potential is available for only a few solvable or weakly
interacting systems.  Furnstahl and
colleagues~\cite{Furn06,Pug03,Bhat05} derived energy-density
functionals for dilute Fermi gases with short-ranged interactions.
For Fermi gases in the unitary regime, simple scaling arguments
suggest the form of the energy-density
functional~\cite{Car03,Pap05,Bhat06,Bul07}.  We similarly use scaling
arguments in the development of our functional.

In recent years, significant progress has been made in developing
global nuclear energy-density functionals; see, for example,
Refs.~\cite{Gor09,Kort10}.  The Skyrme-Hartree-Fock-Bogoliubov mass
functionals by Goriely {\it et al.}~\cite{Gor09} achieve a least-squares
error of about $\chi = 0.58$~MeV with 14 parameters fit to nuclei with
$N,Z\geq 8$.  The 12-parameter UNEDF functionals {\sc UNEDF}nb and
{\sc UNEDF}0 of Ref.~\cite{Kort10} have a least-squares deviation of
$\chi=0.97$~MeV and $\chi=1.46$~MeV, respectively, when fit to a set
of 72 even-even nuclei. The fitted functionals, when applied to 520
even-even nuclei of a mass table, yield least-squares deviations of
$\chi=1.45$~MeV and $\chi=1.61$~MeV, respectively. The
refit~\cite{Bert05} of the Skyrme functional Sly4 to even-even
nuclei exhibits a root-mean-square deviation of about $\chi=1.7$~MeV.
For a recent review on this matter, we refer the reader to
Ref.~\cite{Irina}. The work presented here provides an approach based
on Hohenberg-Kohn DFT, utilizing degrees of freedom that are natural
to the nuclear shell-model, and gives a global description of the
nuclear chart for all nuclei with $N,Z \geq 8$.

We emphasize the difference between an energy functional for nuclear
masses and a mass formula.  Mass formulae, such as the finite-range
droplet model (FRDM)~\cite{Mol95} and Duflo-Zuker mass
formula~\cite{Duf95}, calculate nuclear masses from an assumed nuclear
structure.  The mass formula by Duflo-Zuker, for instance, computes
the binding energy in terms of the shell-model occupations. The
occupations are parameters---not variables---and their values are
taken from the noninteracting shell-model (with modifications due to
deformations).  The form of the Duflo-Zuker mass formula is motivated
by semiclassical scaling arguments, which ensure that the binding
energy scales linearly with the mass number in leading order; see
Ref.~\cite{Zuker2010} for details.  The Duflo-Zuker mass formula
achieves an impressive root-mean-square deviation of $\chi = 0.35$~MeV
with 28 fit parameters. In an occupation number-based energy
functional, the ground-state energy results from a minimization of a
functional, and the occupation numbers are variables that minimize the
functional. In other words, the structure of a nucleus results from
the minimization of a functional and is not assumed. In our
phenomenological construction of an occupation number-based energy
functional, we employ several scaling arguments and terms from the
Duflo-Zuker mass formula.  Recently, the energy functional (in terms
of occupation numbers) was constructed for the pairing
Hamiltonian~\cite{Pap07,Lacroix10,Lacroix11} and the three-level
Lipkin model~\cite{Bert08,Lacroix09}. The analytical insights gained
from these simple models also enter the construction of terms for our
occupation number-based functional.

This paper is organized as follows. In Sect.~\ref{newfunc} we
introduce the form of the energy functional and discuss the relevant
terms. In Sects.~\ref{minimization1} and \ref{minimization2} we
provide a statistical analysis of the functional and a description of
the minimization routines used.  We study the performance of the
functional in calculating nuclear ground-state energies and radii in
Sect.~\ref{results}.  We conclude with a summary in
Sect.~\ref{summary}.

\section{Occupation Number-based Energy Functional}
\label{newfunc}

In this section, we introduce the theoretical foundation for
functionals based on occupation numbers and discuss the terms that
enter the functional.

\subsection{Theoretical Basis}
We first clarify our work within Hohenberg-Kohn DFT~\cite{Hoh64}.
According to Hohenberg and Kohn, for a many-body system with density
$\rho$ in an external potential $v$ there exists a functional of the
density $\mathcal{F_{HK}}(\rho)$ independent of $v$ such that the
ground-state energy is
\begin{equation}
E_{\rm g.s.} = \min
\left\{ \mathcal{F_{HK}}+\int d^3 {\bf r}~v({\bf r}) \rho ({\bf r}) \right\} .
\label{HKTheorem}
\end{equation}
Here the minimization is performed over all one-body densities
$\rho({\bf r})$.

Current nuclear DFT models, such as HFB-17~\cite{Gor09}, are based on
Skyrme forces and employ densities $\rho (\bf{r})$ and currents
depending on spatial coordinates.  The use of such objects seems a
natural choice, particularly in quantum chemistry and condensed matter
systems where one is interested in the electronic structure in the
presence of ions.  For atomic nuclei, however, other representations
might also be interesting or more natural, and we consider functionals
based on occupations of shell-model orbitals.

To present our concept formally, we consider the ground-state energy
$E(\varepsilon_\alpha )$ of an $A$-body system described by a
shell-model Hamiltonian with single-particle energies
$\varepsilon_\alpha$, $\alpha = 1, 2, 3, \ldots$,
\begin{equation}
\sum_\alpha \varepsilon_\alpha a_\alpha^\dagger a_\alpha+\hat{V}  .
\label{shellH}
\end{equation}
Here, $a_\alpha^\dagger$ creates a fermion in the shell-model orbital
$\alpha$, and $\hat{V}$ is the nuclear interaction. The ground-state energy is
a (complicated) function of the single-particle energies, and the
occupation number of the orbital with label $\beta$ is
\begin{equation}
n_\beta \equiv \frac{\partial}{\partial \varepsilon_\beta}E(\varepsilon_{\alpha}) .
\label{shellocc}
\end{equation}
A Legendre transformation of the function $E(\varepsilon_\alpha)$
yields the occupation number-based energy functional~\cite{Lieb83}
\begin{equation}
\mathcal{F} (n_\alpha ) = E( \varepsilon_{\alpha}( n_\beta ) )
-\sum_\alpha n_\alpha \varepsilon_\alpha( n_\beta ) .
\label{genFunc}
\end{equation}
From a technical point of view, the one-body potential---or the
mean-field--- has been employed as an external potential in DFT.
Thus, there is a formal analogy between Hohenberg-Kohn DFT and
occupation number functionals.  The Hohenberg-Kohn theorems remain
valid when we exchange $v(\bf{r}) \leftrightarrow \varepsilon_\alpha$
and $\rho({\bf r}) \leftrightarrow n_\alpha$.  However, the explicit
construction of the occupation number-based energy functional as a
Legendre transform is possible only for exactly solvable
systems~\cite{Pap07,Bert08}. In the following section, we employ
primarily phenomenological arguments for the terms that enter the
functional.

\subsection{Form of the Functional}
\label{form}

The occupation number-based functional is guided by semiclassical
scaling arguments that guarantee nuclear saturation~\cite{Duf95,Zuker2010}.  In
the harmonic oscillator basis, a nucleus of $A$ nucleons has about
$A^{1/3}$ occupied shells below the Fermi level, and about $A^{2/3}$
nucleons are in the highest energetic, completely filled shell. We are
interested in expressing the functional as a sum of simple functions
that depend on occupation numbers. Consider, for example, the energy
from a harmonic mean-field potential for the protons. Let $z_p$ denote
the proton occupation number of the $p$th oscillator shell. The
maximum occupation of the $p$th shell is 
\begin{equation}
pz_p\sim p^3 , \label{eg}
\end{equation}
where $\sim$ indicates ``scales as.''  We use semiclassical arguments
to enforce saturation.  The total energy of the harmonic mean-field 
scales as
\begin{eqnarray}
\sum_p p z_p &\simeq& \int_0^{A^{1/3}}dp~p^3 \\
 &\sim& A^{4/3}  .
\end{eqnarray}
Thus, the multiplication of such an energy term with
 a factor of $A^{-1/3}$ ensures saturation, that is, a scaling of the
energy with $A$ in leading order. The simple arguments based on
filling the major shells of the spherical harmonic oscillator yield 
incorrect shell closures. As a remedy, we follow Duflo and
Zuker~\cite{Duf95} and consider shells where the high-$j$ intruder
orbital of the shell $p+1$ enters the shell $p$. Since a single $j$
orbital of the shell $p$ holds on the order of $p$ nucleons---while a
major shell holds about $p^2$ nucleons---our scaling arguments are
unchanged. In the modified shell-model, the resulting shell closures have 
occupation numbers
\begin{equation}
  2, 6, 14, 28, 50, 82, 126, 184, ...  .
\end{equation}
In what follows, we employ a model space consisting of 15 major
shells.  The shell $p$ has a dimension $d_p=(p+1)(p+2)$, 
(i.e.,  it can
hold up to $d_p$ protons and $d_p$ neutrons). We employ suitable
factors of $d_p$ and $p$ to ensure saturation.

We write the functional as a sum of macroscopic Coulomb and pairing
terms plus a microscopic interaction $\mathcal{F}(n,z)$ that depends
on the occupations $z_p$ of proton shells and $n_p$ of neutron shells
\begin{equation}
  F (c; n, z)= c_c \frac{Z(Z-1)}{A^{1/3}}+c_P \frac{\delta}{\sqrt{A}}+\mathcal{F}(c; n, z)  .
  \label{functional}
\end{equation}
Here, $\delta$ is 1, 0, and -1 for even-even, odd mass, and odd-odd
nuclei, respectively, and $Z$ is the charge number.  
The 17 fit parameters of the full functional are denoted as the
shorthand coefficients 
\begin{equation}
c\equiv \{c_1, c_2, \ldots, c_{11},
c_c,c_P,c_s,c_{as},c_{ss},\tilde{c}_s\} ,
\end{equation}
while
\begin{eqnarray}
z &\equiv& \{ z_1, z_2, \ldots \}  \\
n &\equiv& \{n_1, n_2, \ldots\}
\end{eqnarray}
denote the occupations of proton and neutron shells, respectively.
The microscopic functional $\mathcal{F}(c;n,z)$ has the form
\begin{eqnarray}
\mathcal{F }(c; n, z)&=&\hbar\omega \bigg(V+T_{\rm kin}+I_{\rm 2B} +D\nonumber \\ 
&&+D_{\rm 4B}+\mathcal{T}+M_{\rm 4Bex}+L_{\rm val}\bigg) \nonumber \\
&&+\hbar\tilde{\omega}(\tilde{D}_{\rm 2B}+L+\tilde{L}) .
\label{fullfunc}
\end{eqnarray}
Below, we explain the individual terms entering the functional. 
 The functions $\hbar\omega$ and $\hbar\tilde{\omega}$ are
\begin{eqnarray}
  \hbar \omega(c_s,c_{as},c_{ss}) &=& 1-c_s A^{-1/3}\nonumber \\
  &&-\frac{c_{as}}{1+c_{ss}A^{-1/3}} \frac{T(T+1)}{A^2}, \label{hbarcomplex} \\
  \hbar \tilde{\omega}(\tilde{c}_s) &=& 1+\tilde{c}_s A^{-1/3} , 
  \label{hbarcomplex2}
\end{eqnarray}
and they account for a smooth dependence on the mass and the isospin
$T=N-Z$ (with $N$ being the neutron number).  Each of the terms of the
functional~(\ref{fullfunc}) scales at most as $A$.  Thus, the
functions~(\ref{hbarcomplex}) and~(\ref{hbarcomplex2}) include smooth
surface corrections.  The last term of Eq.~(\ref{hbarcomplex}) provides both a volume- and surface-symmetry correction~\cite{Dan09,Nikolov11}.

We now discuss the individual terms of the functional~(\ref{fullfunc}).
We begin with the volume terms that scale as $A$. These are 
\begin{eqnarray}
  V &\equiv& c_1A , \label{f1}  \\
  T_{\rm kin} &\equiv& c_2A^{-1/3}\sum_p p (z_p+n_p) , \label{f2}  \\
  I_{\rm 2B} &\equiv& c_3A^{-1/3}\sum_p \bigg(\frac{z_p(z_p-1)}{p} \nonumber \\
  && +\frac{n_p(n_p-1)}{p} +\frac{2 n_p z_p}{p}\bigg) , \label{f3}  \\
  D &\equiv& c_4\sum_p\left(
  \frac{\sqrt{d_p}}{2}-\frac{2}{d_p^{3/2}}(z_p-d_p/2)^2 \right)\nonumber \\
  && \times 
  \sum_q \left(\frac{\sqrt{d_q}}{2}-\frac{2}{d_q^{3/2}}(n_q-d_q/2)^2 \right) , \label{f4} \\
  D_{\rm 4B} &\equiv& c_6\sum_p \frac{z_p(d_p-z_p)n_p(d_p-n_p)}{d_p^3} , \label{f6} \\
  L &\equiv& c_7\sum_p \bigg(\sqrt{(z_p+\varepsilon)(n_p+1)}  \nonumber \\ 
  &&+\sqrt{(n_p+\varepsilon)(z_p+1)}\bigg) , \label{f7} \\
  \mathcal{T} &\equiv & c_9\sum_p \sqrt{(n_p-z_p)^2+\varepsilon^2} .
\label{f9}
\end{eqnarray}
Note that $\varepsilon=10^{-3}$ is a regularization parameter in
Eqs.~(\ref{f1})-(\ref{f9}), to ensure differentiability.  The term $V$
of Eq.~(\ref{f1}) is a smooth volume term and accounts for the bulk
energy. The term $T_{\rm kin}$ of Eq.~(\ref{f2}) contains a
contribution of the single-particle kinetic energies. This term
exhibits shell effects and thereby corrects the smooth behavior of the
volume term $V$.  Effects of two-body interactions are captured by the
terms beginning with Eq.~(\ref{f3}). The two-body term $I_{\rm 2B}$ is
motivated by the monopole Hamiltonian in the mass formula by Duflo and
Zuker~\cite{Duf95}. Here, the fermionic nature in the proton-proton
and neutron-neutron interaction is evident.

The term $D$ of Eq.~(\ref{f4}) is technically a two-, three- and four-body interaction and serves to include deformation effects. The form of this term has a linear onset for almost-empty and almost-filled shells and assumes its
maximum at mid-shell with half-filled occupations.  In practice, this
term behaves similar to the function $\min
(n_p,d_p-n_p)$ but does not suffer from discontinuity and 
lack of differentiability.  Similarly, the four-body term $D_{\rm
  4B}$ of Eq.~(\ref{f6}) also accounts for deformation effects.

The term of Eq.~(\ref{f7}) is motivated by the analytical results in
Refs.~\cite{Bert08,Pap07}.  For a two-level system with interactions
between the levels, the minimum energy results in a fractional
occupation of the higher level due to a square root singularity in the
functional.  Such square root terms are present in the functional of
the three-level Lipkin model~\cite{Bert08} and the pairing Hamiltonian
\cite{Pap07,Lacroix10}.  The inclusion of these analytically motivated
terms considerably improves the functional's reproduction of
experimental energies.

The term $\mathcal{T}$ of Eq.~(\ref{f9}) is an isospin term 
and includes the isospin contributions of individual orbitals.  Not
surprisingly, Eq.~(\ref{f9}) turns out to be relevant for the accurate
descriptions of nuclei far from the $N=Z$ line.  For computational
purposes Eq.~(\ref{f9}) is written in square roots rather
than absolute values. For $|n_p-z_p|\gg\varepsilon$ we have
$\sqrt{(n_p-z_p)^2+\varepsilon^2}\simeq|n_p-z_p|$, and this term is
thus practically a differentiable approximation of the absolute value.

The volume terms of Eqs.~(\ref{f1}) to (\ref{f9}) also have surface
corrections due to the discrete sums. However, we also need to
include surface corrections independent of the volume terms.
Technically, surface terms scale as $A^{2/3}$. We employ
\begin{eqnarray}
  \tilde{D}_{\rm 2B} &\equiv& c_5\sum_p
  \left(\frac{z_p(d_p-z_p)}{d_p^{3/2}}+\frac{n_p(d_p-n_p)}{d_p^{3/2}}\right),
  \label{f5} \\
  \tilde{L} &\equiv& c_8A^{-2/3}\sum_p \left(z_p \sqrt{z_p+1}+n_p \sqrt{n_p+1}\right), \label{f8}  \\
  M_{\rm 4Bex} &\equiv& c_{10}\sum_p \frac{1}{p^{23}} \bigg[ (z_p(z_p-1)(z_p-2)(z_p-3))^3 \nonumber \\
  &&+(n_p(n_p-1)(n_p-2)(n_p-3))^3 \bigg], \label{f11}  \\
  L_{\rm val} &\equiv& c_{11}\bigg(\sqrt{(d_f-n_f)n_{f+1}+\varepsilon^2}  
\nonumber \\
  && +\sqrt{(d_f-z_f)z_{f+1}+\varepsilon^2}\bigg). \label{f13} 
\end{eqnarray}
The term $\tilde{D}_{\rm 2B}$ of Eq.~(\ref{f5}) accounts for
microscopic pairing contributions. Its form is motivated by Talmi's
seniority model~\cite{Talmi}, again with a maximum contribution at
mid-shell as with the volume term of Eq.~(\ref{f4}). Likewise, the term
$\tilde{L}$ of Eq.~(\ref{f8}) is a surface correction term to the
volume term~(\ref{f7}).

The term in Eq.~(\ref{f11}) accounts for higher-order effects and
resulted from an efficient scheme to identify relevant contributions
that we describe in Sect.~\ref{relterms}.  The powers of $p$ ensure
the desired $A^{2/3}$ scaling.

The term $L_{\rm val}$ of Eq.~(\ref{f13}) is unique as it involves
only the highest occupied shell ($p=f$) in the noninteracting
shell-model; that is, it treats the shell immediately below (with
occupation $n_f$) and above (with occupation $n_{f+1}$) the Fermi
surface. In the naive shell-model, these shells contain about
$A^{2/3}$ nucleons, and $L_{\rm val}$ thus is a surface term. This
term accounts for energy from particle-hole excitations and turns out
to be useful.  We note that the forms of Eqs.~(\ref{f1})-(\ref{f13}) are symmetric about exchange of protons and neutrons, $z_p \leftrightarrow n_p$.  Therefore, the only isospin breaking term of our functional is the Coulomb term of Eq.~(\ref{functional}).

\subsection{Numerical Implementation}
\label{minimization1}

We consider one of the $N_{\rm pts}$ nuclei of the nuclear chart and
label it by $i$. This nucleus has neutron number $N_i$ and proton
number $Z_i$. The minimization of the functional~(\ref{functional})
with respect to the occupations $n, z$ yields the ground-state energy
\begin{eqnarray}
  E_i(c; n^*(c), z^*(c)) &=& \min_{n,z \in \mathcal{D}_i} F(c;n,z) .
  \label{eq:lower}
\end{eqnarray}
We refer to this minimization as the lower-level optimization. In
Eq.~(\ref{eq:lower}), $n^*(c), z^*(c)$ denote the optimal occupation
numbers (we suppress that they depend on the label $i$) and are taken
from the domain
\begin{eqnarray}
  \lefteqn{\mathcal{D}_i = \bigg\{(n,z): }\\
&& \sum_p n_p =N_i; \sum_pz_p = Z_i; 0\leq n_p,z_p\leq
    d_p \bigg\}. \nonumber 
  \label{eq:lowcon}
\end{eqnarray}
From a mathematical point of view, we deal with an optimization
problem that includes linear equality constraints and bound
constraints.  The functional~(\ref{functional}) is nonlinear but twice
continuously differentiable in the occupation numbers $n,z$ over the
domain $\mathcal{D}$.  Furthermore, given the coefficients $c$, we
have algebraic derivatives of $F$ with respect to $n,z$ and can use
these to solve each lower-level problem. We used the sequential
quadratic programming (SQP) routine {\tt FFSQP} \cite{ffsqp} to solve
the constrained optimization problem (\ref{eq:lower}).

We provided {\tt FFSQP} with analytical expressions for the
gradient of $F$ with respect to $n$ and $z$. These gradients are used 
by {\tt FFSQP} to build second-order approximations of the objective 
while remaining within the constrained domain $\mathcal{D}_i$. 
The value of $F$ is iteratively reduced until further local decreases 
would be infeasible with respect to $\mathcal{D}_i$.  This provides a local minimum in the neighborhood of our starting point at the naive level-filling.  We found that the additional effort of employing the 
derivatives with respect to the occupation numbers provided
significant advantages, in terms of speed and stability, over 
other optimization routines such as {\tt COBYLA}~\cite{Pow94}, 
that do not require derivatives to be available.
  
As a starting point, we provide {\tt FFSQP} with the 
occupation numbers $n$ and $z$ corresponding to the naive
level filling, which are feasible for the domain $\mathcal{D}_i$. 
In practice, a local solution is found within about a dozen function 
and gradient calls. 

The fit coefficients $c$ are then determined by minimizing the sum 
of squared residuals
\begin{equation}
  \chi^2(c) =
  \frac{1}{N_{\rm pts}}\sum_{i=1}^{N_{\rm pts}}
\left[ E_i(c; n^*(c), z^*(c))-E_i^{\rm exp} \right]^2
\label{eq:chi2}
\end{equation}
as a function of $c$. Here, $E_i^{\rm exp}$ denotes the measured
ground-state energy of the nucleus $i$. Technically, we deal with a
multi-level optimization problem because each of the theoretical
energies $E_i$ is a solution of a lower-level optimization of the
functional~(\ref{functional}) over the occupation numbers $n$ and $z$.
Obtaining the occupations from an optimization is in contrast to the
mass formula~\cite{Duf95}, where occupations are fixed and input by
hand.  We refer to the optimization of the coefficients $c$ as the
upper-level optimization problem. This problem is solved with the
program POUNDerS (\underline{P}ractical \underline{O}ptimization
\underline{U}sing \underline{N}o \underline{Der}ivatives for sums of
\underline{S}quares), and details are presented in
Sect.~\ref{minimization2}.

Note the computational complexity present in the minimization of the
chi-square~(\ref{eq:chi2}). For each of the $N_{\rm pts}$ nuclei 
we must perform a lower-level minimization of the
functional to determine occupation numbers that minimize the
functional and yield the ground-state energy $E_i$. The occupation
numbers resulting from this lower-level optimization can be very
sensitive to the coefficients $c$ of the functional, which in turn 
are determined in the upper-level chi-square minimization. To reduce 
the wall clock time of each $\chi^2$ evaluation, we can perform the
lower-level minimizations in parallel, using as many
as $N_{\rm pts}$ processors.  Further details of the $\chi^2$
minimization are presented in Sect.~\ref{minimization2}.

The computation of ground-state energies via the
functional~(\ref{functional}) with coupled minimizations differs
considerably from the uncoupled case.  If the proton and neutron
occupations are kept fixed to the naive level filling,
Eq.~(\ref{functional}) becomes a mass formula.  Performing a
chi-square fit with this fixed filling yields coefficients $c$ that
can then be used when minimizing the functional with respect to
occupations (that is, the lower-level minimization is done
independently of the upper-level minimization).  The resulting error,
$\chi=3.38$~MeV, is one order of magnitude larger than the residual
error of the mass formula by Duflo and Zuker~\cite{Duf95}.  When
including the lower-level optimization in the chi-square fit (and
hence dealing with a multi-level optimization), we obtain a
considerably reduced $\chi=1.31$~MeV.

\subsection{Determination of Relevant Terms}
\label{relterms}

Unfortunately, there is no recipe available for the construction of
the occupation-number-based energy functional, and the reader may
wonder how we arrived at the particular form~(\ref{fullfunc}) of the
functional.  As described in Sect.~\ref{form}, the guiding principles
are saturation, mean-field arguments, and insights from analytically
solvable systems. However, these arguments do not fully constrain the
functional, and we need a more systematic approach to identify terms
that should enter into the functional.

Bertsch {\it et al.}~\cite{Bert05} employed the singular value
decomposition and studied the statistical importance of various linear
combinations of terms that enter the functional. This method
identifies the relative importance of possible combinations of terms
and truncates search directions that are flat in the parameter space.
Along these ideas, we employ a method that chooses new functional
terms based on their correlation to terms already present.  New terms
are chosen to provide a relatively independent ``search direction'' in
the parameter space of the coefficients $c$ in which the $\chi^2$ of
Eq.~(\ref{eq:chi2}) is minimized. This approach is presented in detail
in Subsect.~\ref{corraxtest}.

In mass formulae, the addition of new terms (and new fitting
parameters) yields a chi-square that is a monotone decreasing function
with the number of fit parameters. For a functional, however, matters
are different.  Here, the addition of a new term to the functional
guarantees a lowered chi-square only if the term has a perturbatively
small effect.  This point is discussed in Subsect.~\ref{PerTest}.

\subsubsection{Correlation Test}
\label{corraxtest}

We first describe our method for selecting new terms to be included in
the functional, beyond those strongly motivated by scaling arguments,
mean-field arguments, and solutions to simple analytic systems.  We
seek a systematic method for determining new terms that will provide
further insight into the physical system and decrease the overall 
chi-square. 

Consider the addition of a term $c_f f$, with new fit
parameter $c_f$, to the functional
\begin{eqnarray}
  \lefteqn{F_0(c;n,z) =}\nonumber\\
&&c_c \frac{Z(Z-1)}{A^{1/3}}+c_P
  \frac{\delta}{\sqrt{A}} +\sum_{\alpha=1}^{M} c_\alpha f_\alpha(n^*(c),z^*(c)) .
  \label{func_noiso}
\end{eqnarray}
This functional contains $M$ terms depending on occupation numbers,
$n^*,z^*$ denoting the occupation numbers that minimize $F_0$ for a
given nucleus. The occupation numbers depend on the nucleus $i$ under
consideration, but we suppress this dependency.  One can expect that
the addition of the term $c_f f$ to $F_0$ will be useful in lowering
the chi-square only when it is independent of the $M$ terms already
included in the functional.

For identification of a new search direction we compute 
the correlation coefficient
\begin{eqnarray}
  R_{f_\alpha,f}&=&\frac{\mbox{cov}(f_\alpha,f)}{s_{f_\alpha} s_{f}} . \label{corr}
\end{eqnarray}
Here the covariance is
\begin{equation}
  \mbox{cov}(f_\alpha,f) = \langle f_\alpha f \rangle 
- \langle f_\alpha\rangle \langle f\rangle ,  
\end{equation}
and the average $\langle \cdot \rangle$ is computed with respect to the 
$N_{\rm pts}$ nuclei of the nuclear chart. Similarly, the standard deviations
are 
\begin{equation}
  s_f=\sqrt{\langle f^2\rangle -\langle f\rangle^2} \mbox{ and }
  s_{f_\alpha}=\sqrt{\langle {f_\alpha}^2\rangle -\langle {f_\alpha}\rangle^2} . 
\end{equation}
In the computation of the averages, the terms $f$ and $f_\alpha$ are
evaluated at the occupations $(n^*(c),z^*(c))$ that minimize
Eq.~(\ref{func_noiso}).  Here $(n^*(c),z^*(c))$ depend on the
coefficients $c$ that minimize the $\chi^2$ based on the energies 
in Eq.~(\ref{func_noiso}).
We note that the correlation coefficient is independent of the size of
the coefficient $c_f$ of the new term under consideration, though
dependent on the other fit coefficients $c$ through the optimal
occupations $(n^*(c),z^*(c))$.  Should the correlation be sufficiently
low for all included terms, the new term $c_f f$ becomes a candidate
and is tested for its performance in lowering the $\chi$ value.

This approach allows us to probe many different forms of functional
terms and then scan through several hundred iterations without the
time-consuming and computationally expensive aspects of performing a
full minimization of the chi-square for each new term under question.
In this way, we systematically grow the functional term by term.  We
started from an initial base of about 350 different terms and found
that only 18 of them were weakly correlated to the existing terms and
had the potential to significantly decrease the least-squares error.
Of these 18 terms, 15 were seen to be simply higher-order corrections
of three primary forms.  We further reduced the set of possible terms
through physical arguments and preliminary fits.  This approach showed
Eq.~(\ref{f9}) to be the best choice for lowering our least-squares
deviation.  Equations (\ref{f11}) and (\ref{f13}) were determined
similarly, from very large sets of possible terms.  In this way we
successfully lowered the functional's least-squares error to a
meaningful $\chi = 1.31$ MeV with 17 fit parameters.

\subsubsection{Perturbative Test}
\label{PerTest}

Assume again that our functional $F_0$ is as in Eq.~(\ref{func_noiso}),
and that we consider the addition of a new term $c_f f$ (with the new
fit coefficient $c_f$ taken to be the mean value of the currently
determined fit coefficients, where there are no statistical outliers).
We will consider the general case where the new term $f$ depends also
on the fit coefficients $c$ of $F_0$. In what follows, we consider a
single nucleus with the ground-state energy
\begin{equation}
  E_0 \equiv F_0(c;n^*,z^*)
\label{Egs}
\end{equation}
obtained from the functional $F_0$. Let us assume that
\begin{equation}
f(c;n^*,z^*) \ll E_0.
\end{equation}
Thus, the new term is perturbatively small (assuming that
the new fit coefficient $c_f$ is of ``natural'' order one). Furthermore, we make
the assumption that $f$ is of ``natural'' size and perturbatively small in
a neighborhood of $(n^*,z^*)$ (i.e., its derivatives in this
neighborhood are of the same order as $f$).  We consider
$F=F_0 +c_{\rm new} f_{\rm new}$, and it is clear that the minimum of
$F$ will be found at some new occupations $n^*+\delta n, z^*+\delta z$
with corrections $\delta n$ and $\delta z$ that are much smaller than
$n^*$ and $z^*$, respectively. We expand
\begin{eqnarray}
  \lefteqn{\min_{n,z\in \mathcal{D}_i} \Big(F_0(c;n,z)+c_ff(c;n,z)\Big)}\nonumber \\ 
  &=& 
  F_0(c;n^*+\delta n,z^*+\delta z)+ c_f f(c;n^*+\delta n, z^*+\delta z) \nonumber\\
  &\approx&F_0(c;n^*,z^*) +c_f f(c;n^*,z^*) \nonumber\\
&&+\sum_p\frac{\partial F_0}{\partial n_p}\bigg|_{n^*, z^*}\delta n_p
+\sum_p\frac{\partial F_0}{\partial z_p}\bigg|_{n^*, z^*}\delta z_p \nonumber \\
&&+c_f\sum_p\frac{\partial f}{\partial n_p}\bigg|_{n^*, z^*}\delta n_p
+c_f\sum_p\frac{\partial f}{\partial z_p}\bigg|_{n^*, z^*}\delta z_p\nonumber \\
&\approx& F_0(c;n^*,z^*) +c_f f(c;n^*,z^*), 
 \label{Pertexp}
\end{eqnarray}
and the approximation is due to our limitation to first-order
corrections.  Note that we have expanded to first-order in smallness,
eliminating terms that go as derivatives of $f$ since $|f' \delta n|\ll
|F_0|$ and $|f'\delta p|\ll |F_0|$.  The derivatives of the functional $F_0$ with respect to the occupation numbers vanish at the optimum occupation numbers $(n^*,z^*)$, where it is understood that the variation is only with respect to those
occupation numbers that are not at any of the boundaries (i.e., those
occupation numbers for which $n^*_j\ne 0$ or $d_j$ and $z^*_j\ne 0$ or $d_j$),
and that the variation fulfills the equality constraints. Technically
speaking, these are the {\it reduced} derivatives~\cite{Miel04}.
Thus, in leading order of perturbation theory, the functional is
simply a mass formula (as it is evaluated at the leading-order
occupations), and the chi-square fit cannot yield an increased
root-mean-square error for the ground-state energies. In the worst
case, $c_f=0$ will result from the fit.
Thus, only the addition of terms to the functional that cause
perturbatively small changes to the occupation numbers are expected to
result in a decreased chi-square.  

For quickly evaluating a new candidate term $f$ to be added to
the functional, we begin by treating the full functional as a mass
formula.  That is, we freeze the proton and neutron occupations at the
optimal values determined prior to the addition of the new term and
approximate the ground-state energy in the presence of the new term as
\begin{eqnarray}
  E_i(c;n^*, z^*) \approx F(c;n^*, z^*) + c_f f_{\rm new}(n^*, z^*) .
\label{formula}
\end{eqnarray}
We evaluate the energy~(\ref{formula}) for each nucleus $i$, perform 
a $\chi^2$ minimization of the resulting mass formula~(\ref{formula}),
 and obtain a new set of test coefficients, 
$c_{\rm test}$.  Using these coefficients we recalculate
the occupations $n^*_{\rm test}, z^*_{\rm test}$ and the ground-state energies 
\begin{eqnarray}
  \lefteqn{E_i(c_{\rm test};n^*_{\rm test},z^*_{\rm test}) =} \nonumber \\
  &&\min_{n,z \in \mathcal{D}_i} \Big(F(c_{\rm test};n,z) +c_f f(c_{\rm test};n, z) \Big)
\end{eqnarray}
that minimize the functional.  The computation of the corrections
$(\delta n, \delta z) \approx(n^*_{\rm test}-n^*,z^*_{\rm test}-z^*)$
to the occupation numbers shows whether the new term is perturbative
in character and can thus be expected to lower the chi-square.

\subsection{Model for Nuclear Radii}
\label{radmodel}

In DFT the radius $r$ of a nucleus is computed from the density
$\rho({\bf r})$ as
\begin{equation}
  \langle r^2 \rangle =\frac{1}{A} \int d{\bf r}^3~{\bf r}^2\rho({\bf r}) .
\end{equation}
In constructing the functional we employed a harmonic oscillator basis for 
the shell-model and the scaling arguments. 
Consequently, we also employ it for the computation of $\langle r^2\rangle$.
The expectation value of the radius squared in the harmonic oscillator
shell with principal quantum number $\nu$ and angular momentum
$\lambda$ is
\begin{eqnarray}
\langle \nu \lambda |r^2| \nu \lambda \rangle = \ell^2(2\nu +\lambda+3/2) .
\end{eqnarray}
Here, $\ell$ is the oscillator length and is set to 
$\sqrt{492.5 A^{1/3}}$~fm, and $m$ is the mass of the proton. We use
$p=2\nu+\lambda$ and thus find for the expectation value of the 
charge radius squared  
\begin{equation}
  \langle r^2\rangle= \frac{\ell^2}{Z}\sum_p z_p(p+3/2) .
  \label{radius_p}
\end{equation}
In computing charge radii, we must account for the finite
size of the nucleons~\cite{Friar97},
\begin{equation}
  \langle r_{ch}^2\rangle= \langle r_p^2\rangle+\langle
  r_{ch,p}^2\rangle+\frac{N}{Z}\langle r_{ch,n}^2\rangle. \label{radius_ch}
\end{equation}
Here, the proton and neutron charge radii are $\sqrt{\langle
  r_{ch,p}^2\rangle} = 0.877$~fm, $\langle
r_{ch,n}^2\rangle=-0.1161$~fm$^2$, respectively.

We follow Ref.~\cite{Duflo94} and compute the charge radius within the
following model: 
\begin{gather}
  \mathcal{V}=v_1+v_2r^3+v_3 \frac{N-Z}{r}+v_4\frac{(N-Z)^2}{r^3} \label{Vol} \\
  r_{\rm fit}(v;z)=v_5 \mathcal{V}^{1/3}+v_6 . \label{RadVol}
\end{gather}
Here $r=\sqrt{\langle r_{ch}^3\rangle}$, and the coefficients $v$
parameterize the model.  We perform a $\chi^2$ fit of the radii to
determine the coefficients $v\equiv (v_1,\ldots , v_6)$ in
Eqs.~(\ref{Vol}) and~(\ref{RadVol}), whereby we minimize
\begin{equation}
  \chi_r^2(v) =
  \frac{1}{N_{\rm pts}}\sum_{i=1}^{N_{\rm pts}}\left(r_{{\rm fit},i}(v;z)-r_i^{\rm exp}
  \right)^2 .
\label{eq:chi2Rad}
\end{equation}

 \section{Minimization of the Functional}
\label{minimization2}

The POUNDerS algorithm was developed for nonlinear
least-squares problems where the derivatives of the residuals are
unavailable.  A summary of the algorithm in the context of DFT can be
found in \cite{Kort10}.

For fixed $n$ and $z$, the derivatives $\nabla_c
E_i^{\rm th}(c;n,z)$ are known and continuous except when
$c_{ss}=-A^{1/3}$ for some nucleus; see Eq.~(\ref{hbarcomplex}).
However, the form of the nonlinear lower-level minimization in
Eq.~(\ref{eq:lower}) does not satisfy standard regularity conditions
that would ensure existence and continuity of the derivatives
$\frac{\partial n^*(c)}{\partial c_j}$ and $\frac{\partial
  z^*(c)}{\partial c_j}$ (see, e.g., \cite{Rob80}). Thus,
unavailability of the residual derivatives in our case comes from the
dependence of the optimal occupation numbers $n^*,z^*$ on the
coefficients $c$.

From a theoretical standpoint POUNDerS requires continuously
differentiable residuals. For this functional, however, we found that a
smooth model-based method accounting for the problem structure yielded
better coefficients in fewer simulations than do optimization algorithms
that can explicitly treat the nonsmoothness. A similar result was
found for many of the piecewise smooth problems in \cite{MW09}.

To minimize Eq.~(\ref{eq:chi2}), the POUNDerS used in \cite{Kort10}
was modified to account for known partial derivatives with respect to
some of the coefficients. The optimal occupations
$\{n^*(c)\},\{z^*(c)\}$ are independent of the three coefficients
($c_c,c_P,c_1$) because the corresponding terms in the functional
Eq.~(\ref{fullfunc}) do not involve $n,z$. Hence we can
algebraically compute the (continuous) derivatives
\begin{equation}
  \frac{\partial E_i^{\rm th}}{\partial c_j}, \, \frac{\partial^2 E_i^{\rm th}}{\partial
    c_j\partial c_k}, \qquad j,k\in\{c,P,1\} .
  \label{eq:ders}
\end{equation}

The unavailability of derivatives makes optimization significantly
more challenging. Over the course of the optimization, POUNDerS
effectively builds up coarse approximations to first- and second-order
derivatives of the residuals by interpolating the residuals at
previously evaluated coefficient values.
Knowing the residual derivatives of 3 of the 17 coefficients
effectively lowers the dimension of the difficult derivative-free
optimization problem, resulting in fewer evaluations of $\chi^2$.
\begin{figure}
  \includegraphics[width=0.5\textwidth,angle=0]{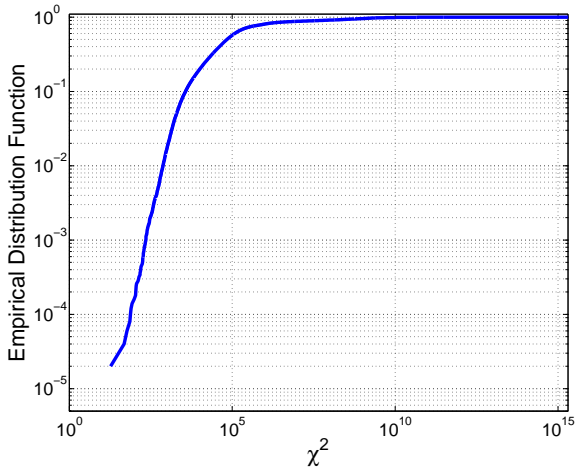}
  \caption{(Color online) $\log-\log$ empirical cumulative distribution
    function showing the probability of randomly finding a value below
    the given $\chi^2$ value in the hypercube $[-1,1]^{17}$.
\label{fig:cdf}}
\end{figure}

To guard against the effects of multiple local $\chi^2$ minimizers and
discontinuities of the computed energies, we found that a sufficient
strategy was periodic restarting of POUNDerS in neighborhoods of mild
size. This allows the local algorithm to occasionally look beyond the
smaller neighborhood it has focused on. While it is impossible to
guarantee that $\chi^2$ has been globally minimized, the local
solution reported in the next section is significantly better than the
$\chi^2$ values found for other coefficients.  Based on $50,000$
$c$ values uniformly drawn from the hypercube $[-1,1]^{17}$, 
Fig.~\ref{fig:cdf} shows that roughly $0.01\%$ of $c$ have
$\chi^2\leq 10^2$ and roughly $50\%$ have $\chi^2\geq 10^5$. This
shows that by taking into account the structure of $\chi^2$ and the
availability of some derivatives, with POUNDerS we were able to find a
proverbial ``needle in a haystack'' with $\chi \approx 1$.

\section{Results}
\label{results}

We now analyze the results of the functional based on coefficients $c$
determined by the $\chi^2$ fit~(\ref{eq:chi2}) of the energy
calculations to experimental data.

\subsection{Energy}
\label{EnPerf}
We fit our functional to a set of 2,049 nuclei from the 2003 atomic
data evaluation~\cite{audi2003} whose uncertainty in the binding
energy is below 200~keV. The resulting fit produces a least-squares
error of $\chi = 1.31$ MeV, and the fit coefficients in units of MeV are 
\begin{eqnarray*}
c_c &=& -0.619948, \\
c_p &=& 11.170908 , \\ 
c_s &=& 0.891816 , \\
c_{as} &=& 7.434098, \\
c_{ss} &=& 0.397623 , \\ 
c_2 &=& 23.174559 , \\
c_3 &=& 0.233345 , \\
c_4 &=& 0.493533 , \\
c_5 &=& -10.678202 , \\
c_6 &=& -0.353447, \\ 
c_1 &=& -7.672829, \\
\tilde{c}_s &=& -0.429029, \\    
c_8 &=& 7.333364 , \\
c_7 &=& 0.112605  , \\
c_9 &=& 0.382828 , \\
c_{10} &=& -4.107110 , \\ 
c_{11} &=& 1.383515 .
\end{eqnarray*}
Our root-mean-square deviation is competitive with
current Skryme force-based functionals~\cite{Kort10}.  While the
current least-squares deviation is greater than the best achieved by a
mean-field based functional~\cite{Gor09}, we provide a novel mass
functional that uses the orbital occupations of nucleons as the
relevant degrees of freedom.  

\begin{figure}
  \includegraphics[width=0.5\textwidth,angle=0]{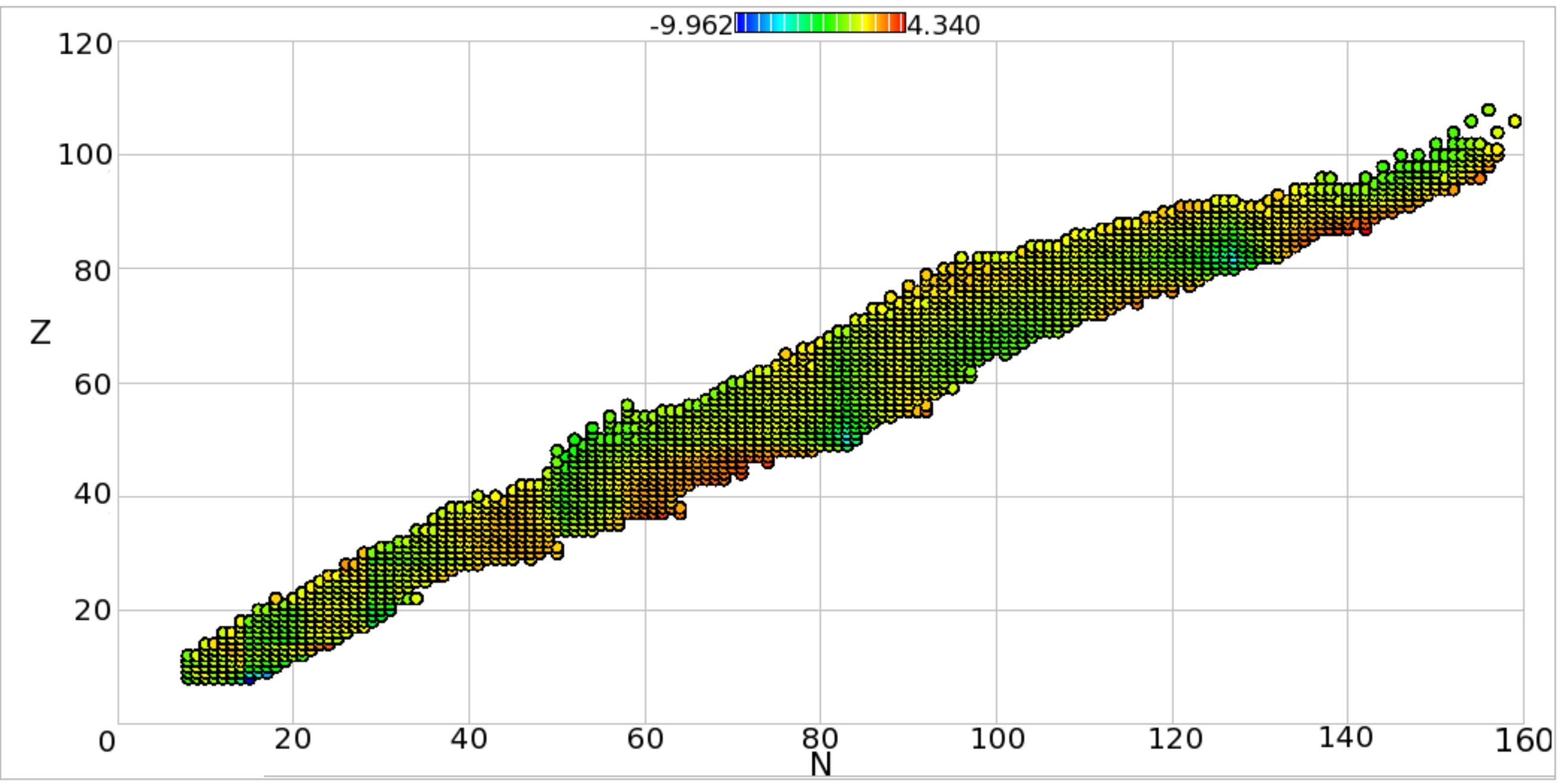}
  \caption{(Color online) Chart of 2,049 well-measured nuclei from the
    2003 atomic data evaluation in Ref.~\cite{audi2003}, 
    color showing the difference $E^{\rm
      th}-E^{\rm exp}$ between the calculated energy and experimental
    energy.  The energy difference exhibits a smooth behavior across
    the whole chart.  Some overbinding is seen in the area of very
    heavy nuclei and tin isotopes.  Color ranges from -9.962~MeV
    (blue) to 4.340~MeV (red).}
 \label{EnDiff1}
\end{figure}

\begin{figure}
  \includegraphics[width=0.5\textwidth,angle=0]{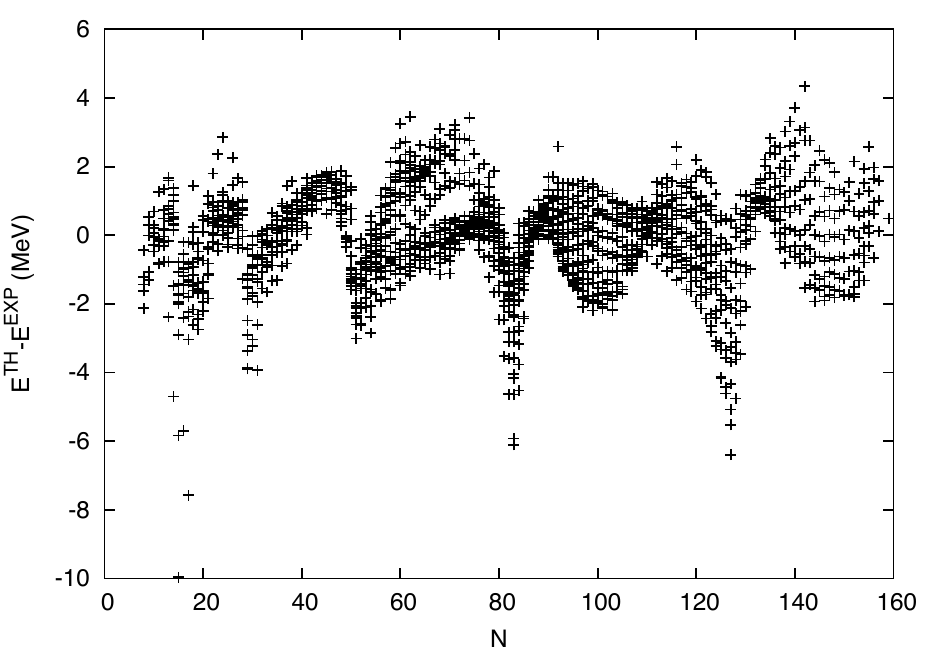}
  \caption{Energy difference $E^{\rm th}-E^{\rm exp}$
    as a function of $N$ for the same nuclei as in Fig.~\ref{EnDiff1}.
    The small oscillations around zero indicate a good description of
    nuclear shell structure.}
    \label{EnN}
\end{figure}

\begin{figure}
  \includegraphics[width=0.5\textwidth,angle=0]{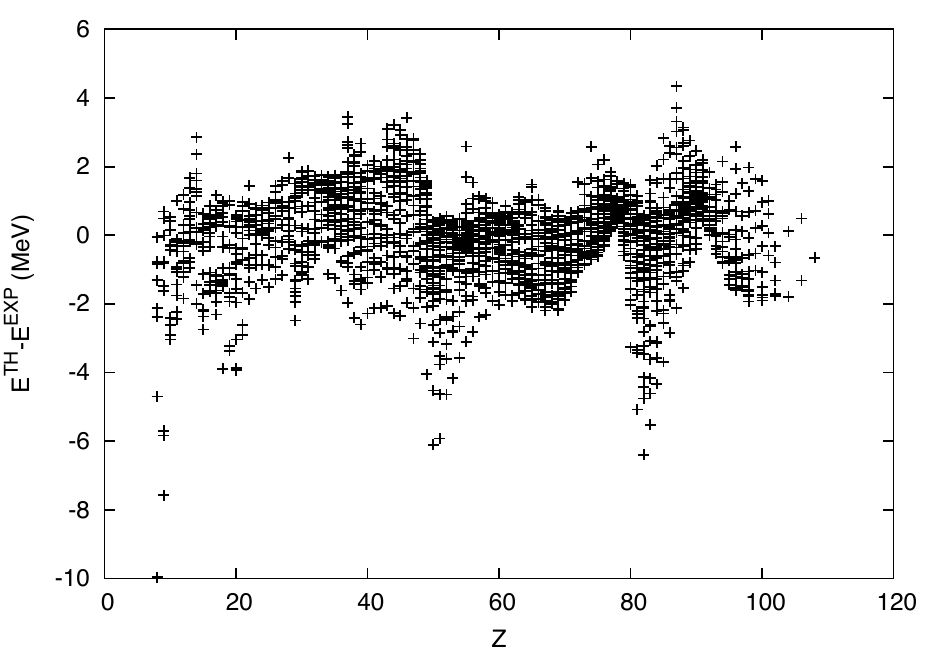}
  \caption{Energy difference $E^{\rm th}-E^{\rm exp}$ as a function of
    $Z$ for the same nuclei as in Fig.~\ref{EnDiff1}.  The small
    oscillations around zero indicate a good description of nuclear
    shell structure.}
  \label{EnZ}
\end{figure}

Figure~\ref{EnDiff1} shows an $N$ vs. $Z$ chart of the fit nuclei 
with color representing the difference $E^{\rm th}-E^{\rm exp}$ 
between the energies computed from the functional and the
experimental data for the 2,049 nuclei employed in the fit. The
differences are a smooth function of the neutron and proton numbers.
Figures~\ref{EnN} and \ref{EnZ} display the energy differences as
functions of $N$ and $Z$.  There are still systematic deviations
associated with shell oscillations. Recall that the shell closures are
input to our functional through the choice of the single-particle
degrees of freedom. Thus, the shell oscillations reflect smaller
deficiencies associated with the description of nuclear deformation.

To test the extrapolation properties of our functional, we fit the
functional to a smaller set of 1,837 nuclei taken from the 1993 atomic
evaluation data set~\cite{audi1993}.  For this data set, the
root-mean-square error is $\chi = 1.38$~MeV. This least-squares
deviation is close to that from a fit to the larger set of 2,049
nuclei, and the deviations are again smooth across the nuclear chart.
Employing the functional from the fit to the 1993 data set, we compute
the ground-state energies of all 2,049 nuclei and find a
root-mean-square deviation of $\chi = 1.34$~MeV.  Thus, the functional
has good extrapolation properties.

\begin{figure}
  \includegraphics[width=0.5\textwidth,angle=0]{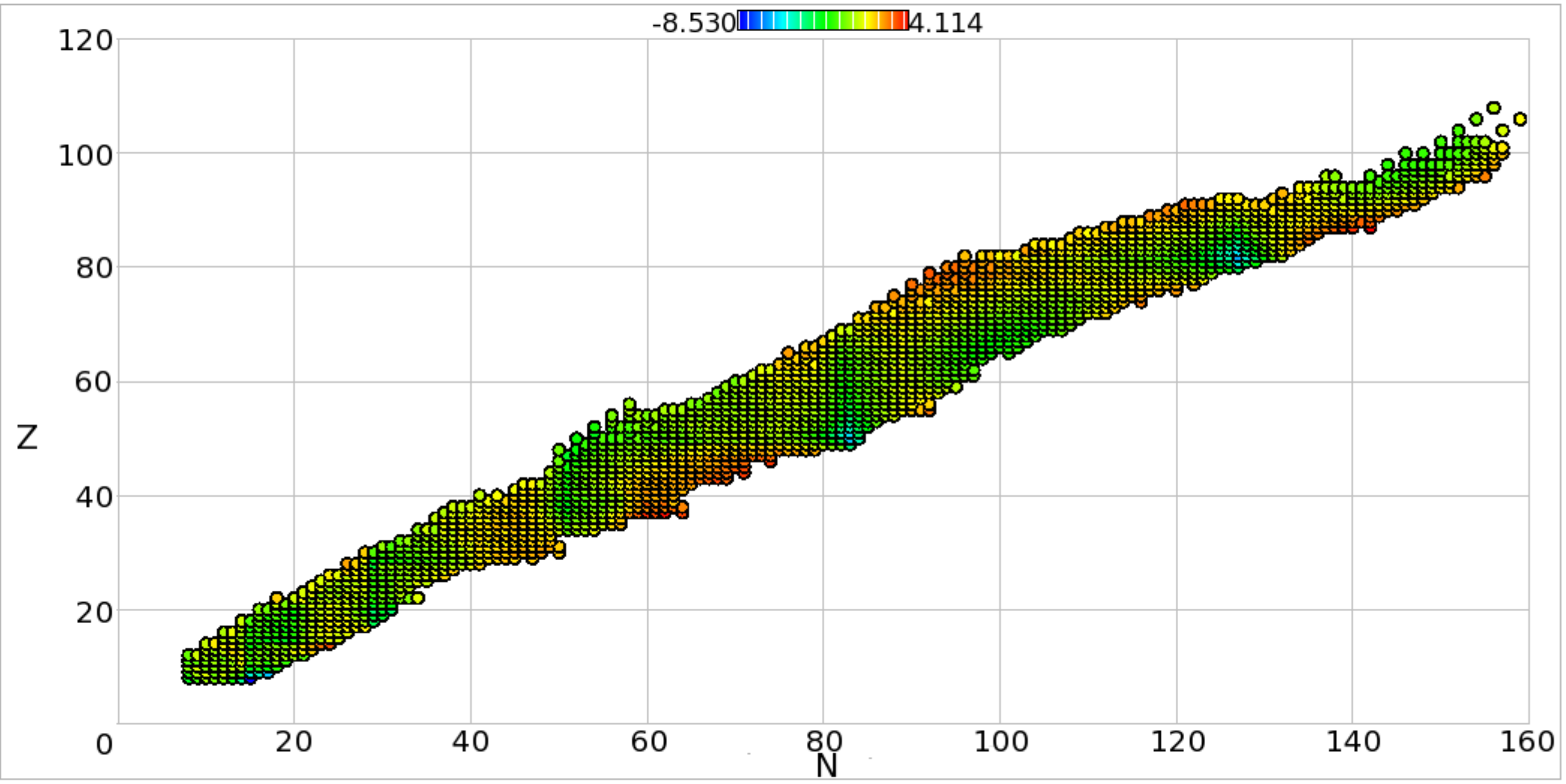}
  \caption{(Color online) Energy difference $E^{\rm th}-E^{\rm exp}$
    between theoretical and experimental results for the functional
    that is fit to the 1993 data set and applied to the subset B of
    well-measured nuclei of the 2003 data set.  The deviations are smooth across the nuclear
    chart and consistent with Fig.~\ref{EnDiff1}.  Color ranges from
    -8.530 MeV (blue) to 4.114 MeV (red).}
    \label{EnDiffex95}
\end{figure}

To further test the functional's predictive power, we use the
coefficients $c$ from the fit to the 1993 data set ($\chi = 1.38$~MeV)
and the coefficients from the fit to the 2,049 nuclei ($\chi =
1.31$~MeV) and compute the binding energies of 2,149 nuclei in the
complete 2003 nuclear data set.  The functional fit to the 1993 data
set yields $\chi = 1.40$~MeV, and the functional from a fit to the
2,049 nuclei yields $\chi = 1.38$~MeV. These values are close to $\chi
= 1.37$~MeV that results from a fit of the functional to the full 2003
data set. Thus, the extrapolation properties of the functional are
quite good.  Table~\ref{ExtrapTable} summarizes the details of how our
functional extrapolates from data set to data set.
Figure~\ref{EnDiffex95} shows the differences between theoretical and
experimental ground-state energies across the chart of nuclei employed
for the extrapolation.

\begin{table}[!ht]
\caption{Root-mean-square deviations of binding energies resulting
  from a global fit of the functional and from extrapolation to larger
  data sets.  Data set A consists of all nuclei of the 1993 mass
  evaluation~\cite{audi1993}, data set C consists of all nuclei of the
  2003 mass evaluation~\cite{audi2003}, and data set B is a subset of
  data set C consisting of well-measured nuclei whose uncertainty in
  the mass is below 200~keV. The number of nuclei in each data set is
  denoted by $N_{\rm pts}$.\label{ExtrapTable}}
\begin{tabular}{c| c| c| c| c}
  \hline
  Data && \multicolumn{3}{c}{$\chi $ (MeV)} \\
  \cline{3-5}
  Set &&& \multicolumn{2}{c}{Extrapolation to} \\
  \cline{4-5}
  &$N_{\rm pts}$&Fit&Data Set B & Data Set C \\
  \hline
  A&1837& 1.38 &1.34&1.40\\
  B&2049&1.31 & -- & 1.38\\
  C&2149& 1.37 & -- & --\\
  \hline
\end{tabular}
\end{table}

\begin{figure}
  \includegraphics[width=0.5\textwidth,angle=0]{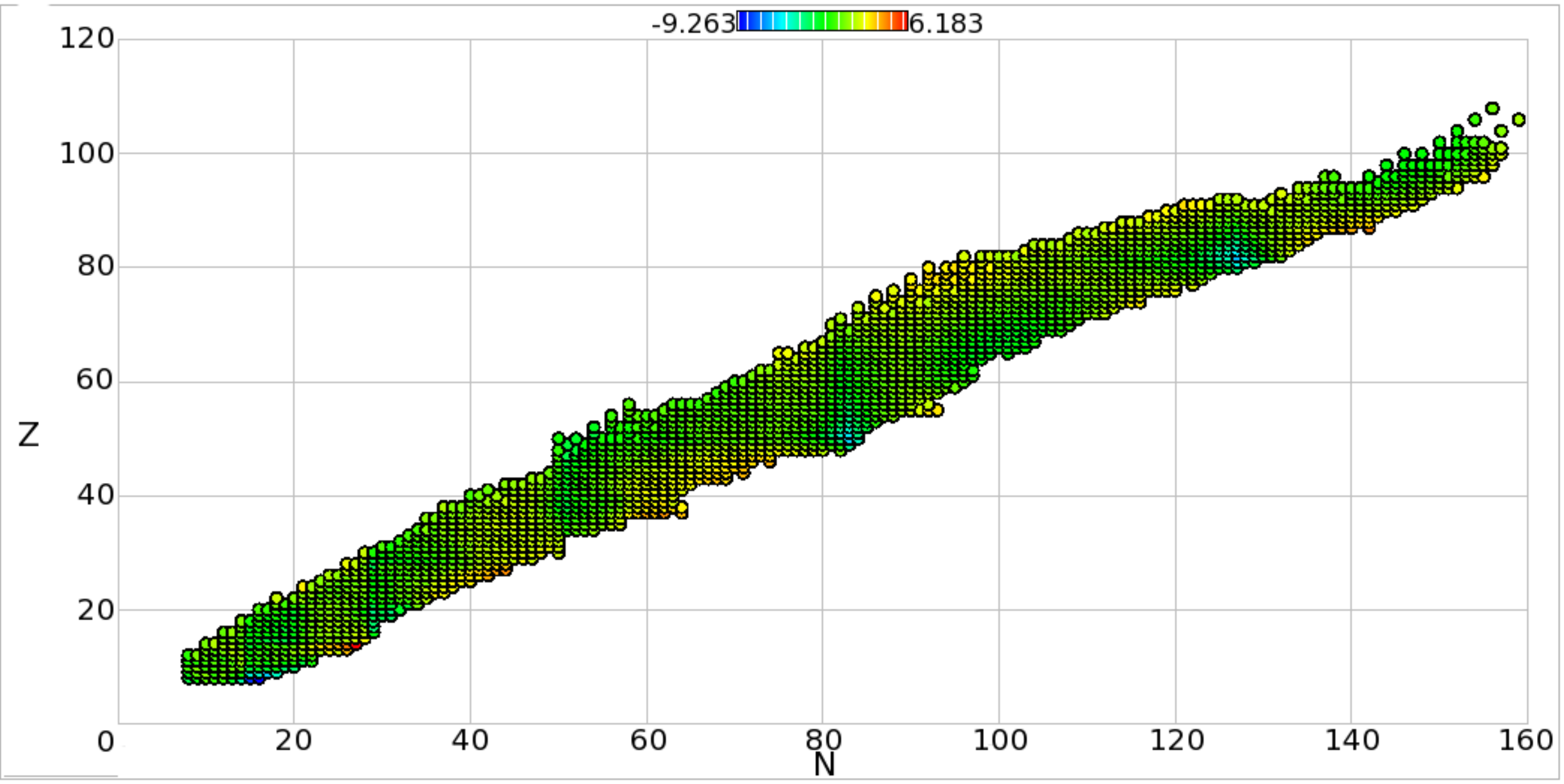}
  \caption{(Color online) Energy difference $E^{\rm th}-E^{\rm exp}$
    between theoretical and experimental results for the functional
    fit to the 1993 data set and applied to the 2003 data set.  The
    deviations are smooth across the nuclear chart and consistent with
    Fig.~\ref{EnDiff1}.  Color ranges from -9.263 MeV (blue) to 6.183
    MeV (red).}
  \label{EnDiffex03}
\end{figure}

\subsection{Radii}
\label{RadPerf}

\begin{figure}[h]
  \includegraphics[width=0.5\textwidth,angle=0]{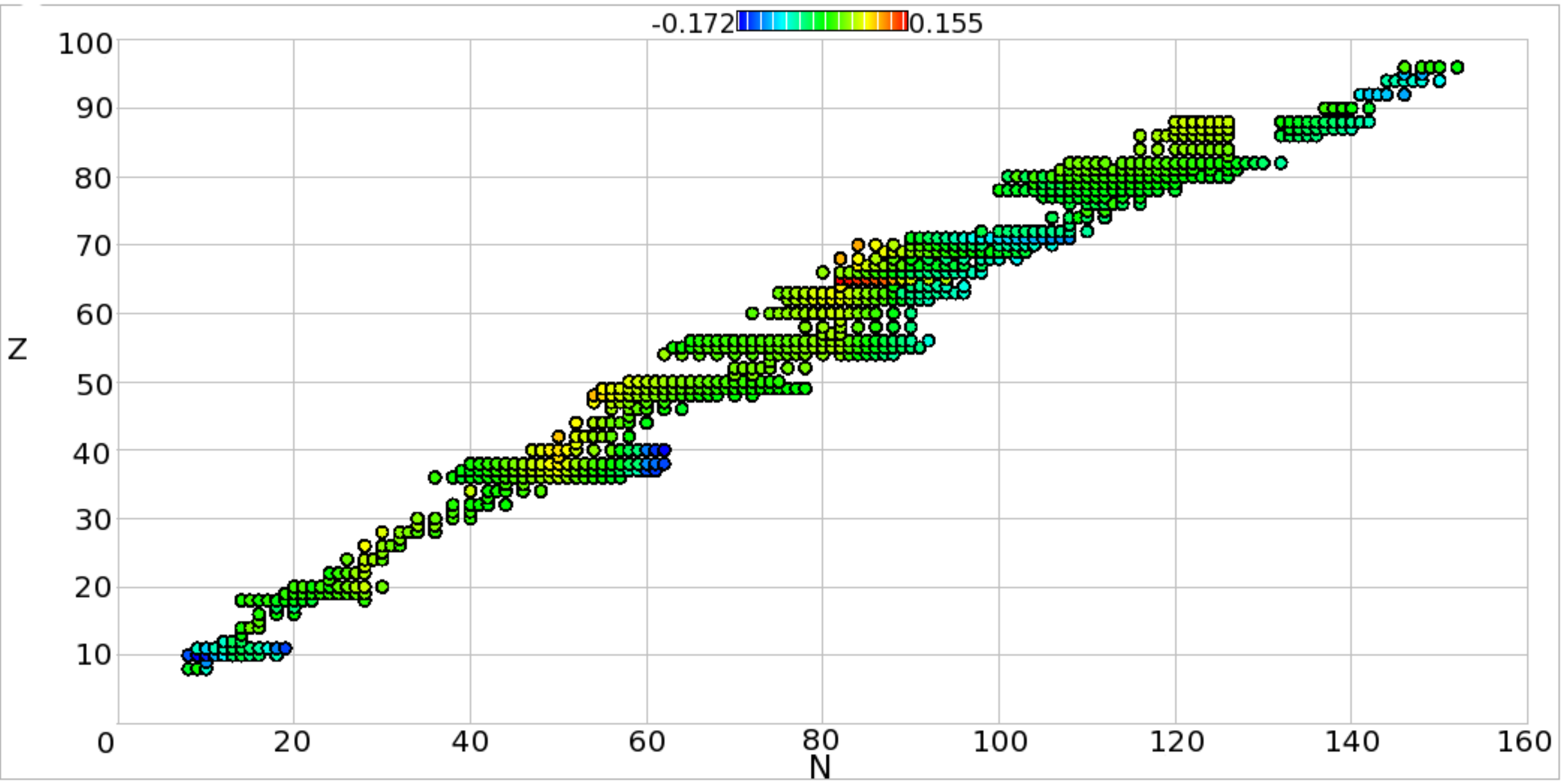}
  \caption{(Color online) Difference between experimental and
    theoretical charge radii for the set of 772 nuclei from
    Ref.~\cite{Angeli04}. Color ranges from -0.172 fm (blue) to 0.155
    fm (red).}
  \label{RadDiff}
\end{figure}

\begin{figure}
  \includegraphics[width=0.75\textwidth,angle=0]{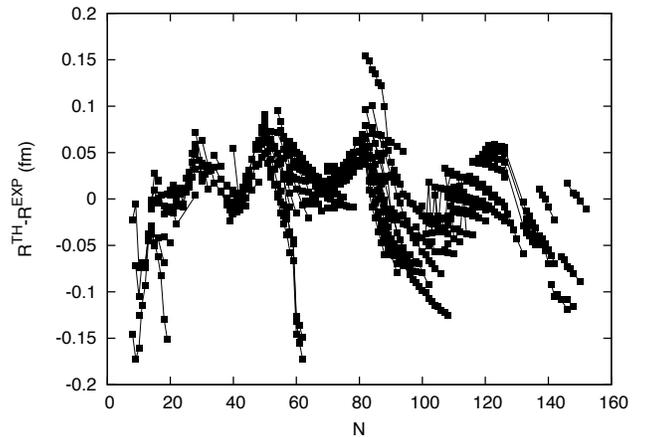}
  \caption{Chart of difference between calculated and experimental charge radii for 772 nuclei as a function of $N$.  Lines connect isotopes.
  \label{RadDiff2}}
\end{figure}

We now turn to the results for nuclear charge radii.  We fit our
six-parameter model~(\ref{RadVol}) for the charge radii to the
experimental data of 772 nuclei from Ref.~\cite{Angeli04} and obtain a
least-squares deviation of $\chi_r = 0.047$~fm. The resulting fit
coefficients (with units of fm) are
\begin{eqnarray}
v_1 &=& 11.004005 , \nonumber\\      
v_2 &=& 2.000870 , \nonumber\\
v_3 &=& -3.248108 , \nonumber\\
v_4 &=& -0.279495 , \nonumber\\
v_5 &=& 0.775617, \nonumber\\
v_6 &=& -0.557746. \nonumber
\end{eqnarray}
Note that the data set contains both spherical and deformed nuclei.
Figure~\ref{RadDiff} shows the difference between the calculated and
experimental radii. For a comparison, note that Duflo and Zuker state
a least-squares error of about $\chi_r \simeq 0.01$ fm for the charge
radii of spherical nuclei \cite{Duflo94}. Radii from Skyrme
functionals exhibit a root-mean-square deviation of about
0.025~fm~\cite{Irina}.  Figure~\ref{RadDiff2} shows the difference
between computed and experimental charge radii as a function of
neutron number.  The outliers seen in Fig.~\ref{RadDiff2} correspond to isotopic chain of Tb and the elements Rb, Sr and Zr with neutrons $N=60-62$.

Let us also study the extrapolation properties of our mass model.  We
fit the model~(\ref{RadVol}) to the charge radii on a subset of 494
nuclei (taken from Ref.~\cite{Angeli04}) chosen near the valley of
stability. This yields a least-squares error of $\chi_r = 0.046$~fm.
When using this model to compute the charge radii on
the full set of 772 charge radii, we find a slightly increased
$\chi_{r} =0.048$~fm, indicating that the extrapolation is successful.

For each nucleus, the occupation numbers enter the computation of the
charge radius.  We thus need to understand how our model for the
charge radius depends on the functional employed for the computation
of binding energies; in other words, its dependence on the
coefficients $c$.  A sensitivity analysis of our fit to binding
energies provides us with a confidence interval for each of the
coefficients $c$.  We take a randomly chosen sample of five sets of
coefficients $\{c_1,\dots,c_5\}$ within the confidence interval and
recompute the structure (i.e., the occupation numbers) and the binding
energies.  The resulting least-squares deviations for binding energy
range from $\chi = 1.34$~MeV to $1.78$~MeV. Subsequently, we compute
the charge radii for the nuclei of interest (without refitting the
coefficients $v$ of our model for the radii). If we refit the
coefficients $v$ of our mass model and adjust them to the change in
the coefficients $c$ of the energy functional, the least-squares error
for the radii changes by at most 4\%. Thus, we find the surprising
result that the model for radii is relatively independent of the
functional's fit coefficients $c$.

\section{Summary}
\label{summary}

We constructed an occupation number-based energy functional for the
calculation of nuclear binding energies across the nuclear chart.  The
relevant degrees of freedom for the functional are the proton and
neutron orbital occupations in the shell-model. A global fit of a
17-parameter functional to nuclear masses yields a least-squares
deviation of $\chi = 1.31$~MeV for the binding energies, and a simple
six-parameter model for the charge radii yields a root-mean-square
deviation of $\chi_r = 0.047$~fm.  The functional has good
extrapolation properties, evident from the application of the
functional fit to data from the 1993 atomic mass evaluation to the
nuclei of the 2003 atomic mass evaluation.  The form of the functional
is guided by scaling arguments and by results from analytically
solvable models.  Isospin and surface correction terms in the
functional proved to be important.  These terms were determined
by a systematic investigation of correlations among possible terms
for the functional and by an analysis of their perturbative behavior.
Additional terms, and therefore lower least-squares deviations, may be
obtained through further use of this method, as well as greater
investigation of the microscopic contributions of higher-order effects
in surface and radial terms.

\begin{acknowledgments}
  We acknowledge communications and discussions with G.~F.~Bertsch, M.~Kortelainen, J.~Mor\'{e}, and A.~P.~Zuker.  This research was supported in part by the U.S. Department of Energy
  under Contract Nos.\ DE-FG02-96ER40963, DE-FG02-07ER41529
  (University of Tennessee), with UT-Battelle,
  LLC (Oak Ridge National Laboratory); the Office of Advanced
  Scientific Computing Research, Office of Science, U.S. Department of
  Energy, under Contract DE-AC02-06CH11357 (Argonne National
  Laboratory); the Office of Nuclear Physics, US Department of Energy
  under Contract DE-FC02-09ER41583 (UNEDF SciDAC collaboration); and
  the Alexander von Humboldt Foundation.  Computing resources were
  provided by the Laboratory Computing Resource Center at Argonne
  National Laboratory.
\end{acknowledgments}

\bibliography{MassBib}

\end{document}